\documentclass[aip,pop]{revtex4-1}
\usepackage{bm,amsmath,amssymb}
\usepackage{graphicx}
\usepackage{subfigure}

\newcommand{\vzero}{{\bf 0}}
\newcommand{\td}{{\rm d}}
\newcommand{\pd}{\partial}
\newcommand{\nab}{\nabla}
\newcommand{\pr}{\prime}
\newcommand{\hrvz}{\hat{\bm z}}
\newcommand{\muo}{\mu_{0}}
\newcommand{\vx}{{\bm x}}
\newcommand{\vu}{{\bm u}}
\newcommand{\vv}{{\bm v}}
\newcommand{\vB}{{\bm B}}
\newcommand{\veps}{\varepsilon}
\newcommand{\vphi}{\varphi}
\newcommand{\betao}{\beta_{0}}
\newcommand{\vA}{v_{\rm A}}
\newcommand{\tauA}{\tau_{\rm A}}

\usepackage{mathtools}
\usepackage{xparse}
\makeatletter
\newcommand{\raisemath}[1]{\mathpalette{\raisem@th{#1}}}
\newcommand{\raisem@th}[3]{\raisebox{#1}{$#2#3$}}
\makeatother
\NewDocumentCommand{\iotabar}{O{0pt} O{0pt}}{% \newhbar[<horz len>][<vert len>]
  \ensuremath{\mathrlap{\raisemath{#2}{\hspace*{#1}{\mathchar'26\mkern-9mu}}}\iota}%
}
\newcommand{\biota}{\iotabar[-0.08em][-0.6ex]}

\draft % marks overfull lines with a black rule on the right

\begin{document}

%\preprint{}

\title{
Calculation of large-aspect-ratio tokamak and
toroidally-averaged stellarator equilibria of high-beta reduced
magnetohydrodynamics via simulated annealing
} %Title of paper

\author{M. Furukawa}
\email[]{furukawa@tottori-u.ac.jp}
\author{Takahiro Watanabe}
%\thanks{}
\affiliation{
Grad. Sch. Eng., Tottori Univ., 
Minami 4-101, Koyama-cho, Tottori-shi, Tottori 680-8552, Japan
}

%\author{Takahiro Watanabe}
%%\email[]{Your e-mail address}
%%\homepage[]{Your web page}
%%\thanks{}
%%\altaffiliation{}
%\affiliation{
%Grad. Sch. Eng., Tottori Univ., 
%Minami 4-101, Koyama-cho, Tottori-shi, Tottori 680-8552, Japan
%}

\author{P. J. Morrison}
%\thanks{}
\affiliation{
Phys. Dept. and Inst. Fusion Studies, Univ. Texas at Austin, 
TX, 78712, USA
}

\author{K. Ichiguchi}
%\thanks{}
\affiliation{
National Institute for Fusion Science, 
322-6 Oroshi-cho, Toki 509-5292, Japan
}
\affiliation{
SOKENDAI, The Graduate University for Advanced Studies, Toki, 509-5292,
Japan
}


\date{\today}

\begin{abstract}
A simulated annealing (SA)  relaxation  method is used  for
 calculation of  high-beta reduced magnetohydrodynamics (MHD) equilibria
 in toroidal geometry. The SA method, based on artificial dynamics
 derived from the MHD Hamiltonian structure,  is used to calculate
 equilibria of large-aspect-ratio and circular-cross-section tokamaks as
 well as toroidally averaged stellarators.  Tokamak equilibria including
 incompressible poloidal rotations are obtained and  the   Shafranov
 shift is seen to  increase  nearly quadratically in the rotation speed.
A mapping procedure between non-rotating and poloidally rotating equilibria
is shown to  explain a quadratic dependence of equilibria shift on rotation.
Calculated stellarator equilibria are seen to agree reasonably with
 previous results.  The numerical results demonstrate the first successful application of
 the SA method to obtain toroidal equilibria.
\end{abstract}


\maketitle %\maketitle must follow title, authors, abstract and \pacs

%\tableofcontents
\section{Introduction}
\label{sec:introduction}

The calculation  of magnetohydrodynamics (MHD) equilibria is an  indispensable
element of basic plasma physics,  as well as the application to magnetic fusion research.
Many methods for  equilibrium calculations have been developed for  two-dimensional and three-dimensional toroidal plasmas.
The present work utilizes and develops the  alternative  simulated
annealing (SA) method of Ref.~\onlinecite{Flierl-Morrison-11} for the
calculation of equilibria.  This method is based on the fact that ideal
fluid dynamics, including ideal MHD,  possesses a Hamiltonian form in
terms of a noncanonical  Poisson bracket.\cite{Morrison-80}  Because of
the Hamiltonian form, ideal  MHD conserves the energy of the system and
has so-called Casimir invariants that arise from degeneracy of the
Poisson bracket.  With the  SA method we solve an artificial dynamics
that is derived so that the energy of the system changes monotonically
while the Casimir invariants are preserved.  Given that equilibria  are
obtained by extremizing the energy at fixed Casimir
invariants,\cite{KO-58, Arnold-65-1, Morrison-98} we can obtain an MHD equilibrium by solving the artificial SA 
dynamics as an initial-value problem. 

A SA method was first developed for two-dimensional neutral fluid flows.\cite{Vallis-89, Carnevale-90,
Shepherd-90}   The method was significantly generalized in Ref.~\onlinecite{Flierl-Morrison-11} by developing several kinds of
brackets that monotonically change the energy of the system while  preserving the Casimir or other invariants of the original system.  A symmetric 
 bracket with the possibility of  various effects such as smoothing through a symmetric kernel and  the preservation of chosen constraints, ones  not inherent in the original dynamics,  by making use of  a Dirac bracket was introduced,  and the general use of such brackets for obtaining equilibria was termed  ``simulated annealing.''

The SA method was applied to low-beta reduced MHD\cite{Strauss-76} in  Ref.~\onlinecite{Chikasue-15-PoP}, using the reduced MHD Hamiltonian structure of   Refs.~\onlinecite{Morrison-84} and \onlinecite{Marsden-84}.  Various equilibria were successfully obtained in a two-dimensional rectangular domain with periodic boundaries in both dimensions.  Moreover, a method was developed for pre-adjusting values of the Casimir invariants of an initial condition for the SA.\cite{Chikasue-15-JFM}   More recently, low-beta reduced MHD equilibria in  cylindrical geometry were  calculated by  SA, where the inside of the plasma was helically deformed, even including magnetic islands.\cite{Furukawa-17}
The symmetric bracket with a smoothing effect was adopted in these studies, demonstrating the  usefulness of  SA for the MHD equilibrium calculation.  For an overview of structure and structure-preserving algorithms for plasma physics from much wider view point, including SA, we refer  to Ref.~\onlinecite{Morrison-17}. 

In the present paper  SA is applied to calculate high-beta reduced MHD\cite{Strauss-77} equilibria in  toroidal geometry, which is possible because 
 high-beta reduced MHD is also a Hamiltonian system.\cite{Morrison-84}  Large-aspect-ratio and circular-cross-section tokamaks as well
as toroidally averaged stellarator equilibria are calculated.  The results obtained  take into account the effect of toroidicity,
where a proper accounting of the  Shafranov shift\cite{Shafranov-58} is observed.    
We compare our  numerical results with previous studies, and obtain
reasonable agreement.   For tokamak equilibria we also include
 incompressible poloidal rotation.  This highlights one of the
 advantages of the SA method.   
The results given here comprise a
 significant next step  towards  ultimate goal of calculating MHD
 equilibrium in fully toroidal geometry.  

Furthermore,  in the present paper we use  a mapping procedure between non-rotating and poloidally
 rotating equilibria for high-beta reduced MHD.  This is an extension to high-beta reduced MHD of the low-beta reduced MHD map first given in Ref.~\onlinecite{Morrison-86}, which is a special case of the generalization of Refs.~\onlinecite{Throumoulopoulos-99} and \onlinecite{Andreussi-12}.  This map nicely explains how equilibria change with poloidal rotation.

This paper is organized as follows.  In Sec.~\ref{sec:formulation}, the
theory of the SA is described for  high-beta reduced MHD.  
Next we will explain numerical schemes for solving the equations of the
SA and the convergence criterion in Sec.~\ref{sec:scheme-convergence}.
Then equilibria of  tokamaks and  toroidally averaged stellarators are
presented in Sec.~\ref{sec:results}. 
The results include tokamak equilibria with poloidal rotation.
There these numerical results are compared and contrasted with previous
studies.  The mapping procedure is also presented in Sec.~\ref{sec:results}. 
Conclusions are given in Sec.~\ref{sec:conclusions}.

\section{Formulation}
\label{sec:formulation}

%In this section, we present the formulation of the SA
%for the high-beta reduced MHD.\cite{Strauss-77} 

\subsection{High-beta reduced MHD and normalization}
\label{subsec:high-beta-RMHD}

The high-beta reduced MHD equations\cite{Strauss-77} are given by
\begin{alignat}{2}
  \frac{\pd U}{\pd t} 
&= 
 [ U, \vphi ] + [ \psi, J ] - \veps \frac{\pd J}{\pd \zeta}
 + [ P, h ]
&\,=:\,&
%f^{U},
f^{1},
\label{eq:vorticity-eq}
\\
  \frac{\pd \psi}{\pd t} 
&=
 [ \psi, \vphi ] - \veps \frac{\pd \vphi}{\pd \zeta}
&\,=:\,&
%f^{\psi},
f^{2},
\label{eq:Ohm-law}
\\ 
 \frac{\pd P}{\pd t} 
&= 
 [ P, \vphi ]
&\,=:\,&
%f^{P}.
f^{3}.
\label{eq:pressure-eq}
\end{alignat}
Here, 
the fluid velocity $\vv$ and the magnetic field $\vB$ are expressed by
\begin{align}
 & \vv = \hrvz \times \nab\vphi, 
\\
 & \vB = \nab\psi \times \hrvz  + \hrvz,
\end{align}
where the unit vector in the toroidal direction is denoted by $\hrvz$.
A coordinate system $(r, \theta, \zeta)$ is used, where 
$r$ is the minor radius, $\theta$ is the poloidal angle and $\zeta$ is
the toroidal angle.  
The length in the toroidal direction is measured by
$z := R_{0} \zeta$ with $R_{0}$ being the major radius of the toroidal
plasma and  $x := r \cos \theta$.
The stream function and the magnetic flux function are denoted by 
$\vphi$ and $\psi$, respectively, with the vorticity in the $\zeta$-direction defined by 
$U := \bigtriangleup_{\perp} \vphi$ 
and the current density in the negative $\zeta$-direction defined by
$J := \bigtriangleup_{\perp} \psi$.
The two-dimensional gradient operator $\nab_{\perp}$ is defined in the
$r$--$\theta$ plane, and the corresponding Laplacian is defined as usual by 
$\bigtriangleup_{\perp} := \nab_{\perp} \cdot \nab_{\perp}$.
The Poisson bracket between any functions $f$ and $g$ is  defined by
$[f, g] := \hrvz \cdot \nab f \times \nab g$.
Quantities appearing in the equations above are normalized by
the plasma minor radius $a$,
a typical mass density $\rho_{0}$,
a typical pressure $p_{0}$,
a toroidal magnetic field $B_{0}$,
an Alfv\'en velocity defined by $\vA := {B_{0}}/{\sqrt{\muo \rho_{0}}}$
where $\muo$ is the vacuum permeability,
and an Alfv\'en time defined by $\tauA := {a}/{\vA}$.
An inverse aspect ratio is defined by $\veps := {a}/{R_{0}}$.
Finally, $P := \betao p$ with $\betao := {2 \muo p_{0}}/{B_{0}^{2}}$ 
and $p$ being the normalized pressure,
and $h := \veps x$.
The right-hand sides 
%$f^{U}$, $f^{\psi}$ and $f^{P}$ 
$f^{1}$, $f^{2}$ and $f^{3}$ 
are defined for
later use.

Note that if the $\zeta$-derivative terms in Eqs.~(\ref{eq:vorticity-eq})
and (\ref{eq:Ohm-law}) are dropped  then the system reduces to  axisymmetric
dynamics.  The effect of the toroidal geometry appears only
through the last term of Eq.~(\ref{eq:vorticity-eq}).

We also note that the reduced MHD equations 
(\ref{eq:vorticity-eq})--(\ref{eq:pressure-eq}) have the same form 
as the reduced MHD equations for helical plasmas 
derived under the stellarator expansion.\cite{Wakatani-Stellarator}   However,  three changes of
Eqs.~(\ref{eq:vorticity-eq})--(\ref{eq:pressure-eq}) are needed for stellarators. First, the toroidal angle $\zeta$ is understood as a toroidal angle for  
long wavelength structures; short wavelength structures are averaged out
in the toroidal direction.
Second, the poloidal flux function $\psi$ is replaced by a total
poloidal flux function
$\Psi := \Psi_{\rm h} + \psi$, where $\Psi_{\rm h}$ is a helical flux
generated by external coils.
Third, the curvature term $h$ is replaced by $\frac{1}{2}\Omega$, where
$\Omega$ is the sum of the curvature of the toroidal magnetic field
and that of the helical magnetic field.
Details of   $\Psi_{\rm h}$ and $\Omega$ will be given
in a following section.

\subsection{Hamiltonian formulation}
\label{subsec:Hamiltonian}

For the high-beta reduced MHD equations the Hamiltonian form was presented in Refs.~\onlinecite{Morrison-84} and \onlinecite{Marsden-84}.  The Hamiltonian is given by 
\begin{equation}
 H[ \vu ]
:=
 \int_{\cal D} \td^{3}x \,
      \left(
         \frac{1}{2} \left| \nab_{\perp} ( \bigtriangleup_{\perp}^{-1} U )
	             \right|^{2}
       + \frac{1}{2} \left| \nab_{\perp} \psi \right|^{2}
       - h P
      \right),
 \label{eq:Hamiltonian}
\end{equation}
where the state vector $\vu := ( U, \psi, P )^{\rm T}$ and the domain is denoted by ${\cal D}$.

The functional derivative of $H[\vu]$ is defined through
\begin{equation}
 \lim_{\delta \vu \rightarrow \vzero} ( H[\vu+\delta \vu] - H[\vu] )
=
 \int_{\cal D} \td^{3}x \, 
 \frac{\delta H[\vu]}{\delta u^{i}} \delta u^{i},
\end{equation}
where $i=1,2,3$ and we sum  over the repeated indices.   Then ${\delta H[\vu]}/{\delta \vu}$ is obtained as
\begin{equation}
 \frac{\delta H[\vu]}{\delta \vu}
=
 \left(
  \begin{array}{c}
   -\vphi
   \\
   -J
   \\
   -h
  \end{array}
 \right).
\label{eq:delHdelu}
\end{equation}
By defining the Poisson operator as
\begin{equation}
 {\cal J}
:=
 \left(
  \begin{array}{ccc}
   -[ U, \circ ] 
   & 
   -[ \psi, \circ ] + \veps \frac{\pd}{\pd \zeta}
   & 
   -[ P, \circ ]
   \\
   -[ \psi, \circ ] + \veps \frac{\pd}{\pd \zeta}
   & 
   0 
   & 
   0
   \\
   -[ P, \circ ]
   &
   0 
   & 
   0
  \end{array}
 \right),
 \label{Jten}
\end{equation}
high-beta reduced MHD can be written  in   Hamiltonian form as
\begin{equation}
 \frac{\pd \vu}{\pd t} 
=
 {\cal J} \frac{\delta H[\vu]}{\delta \vu}=\{\vu, H\}\,,
\end{equation}
where in the last equality the Poisson bracket is used.  A general noncanonical Poisson bracket has the form   
\begin{equation}
 \{ F, G \}
:=
 \int_{\cal D} \td^{3}x^{\pr}
 \int_{\cal D} \td^{3}x^{\pr\pr} \,
 \frac{\delta F[\vu]}{\delta u^{i}(\vx^{\pr})}
 {J}^{ij}( \vx^{\pr}, \vx^{\pr\pr} )
 \frac{\delta G[\vu]}{\delta u^{j}(\vx^{\pr\pr})},
\label{eq:Poisson-bracket}
\end{equation}
where  ${J}^{ij}$ denoting the components of a Poisson operator.  In our case $J=\delta(\vx'-\vx'')\mathcal{J}$, with $\mathcal{J}$   given by Eq.~\eqref{Jten}. (See Ref.~\onlinecite{Morrison-98} for further details.)

\subsection{Casimir invariant}
\label{subsec:Casimir}

A Casimir invariant is defined to be  a functional $C[\vu]$ that satisfies 
$\{ C, F \} = 0$ for any functional $F[\vu]$, where the curly bracket
denotes a noncanonical (degenerate) Poisson bracket.  It is easily shown that the  high-beta reduced MHD bracket defined by \eqref{Jten} has the following three Casimir invariants:
\begin{align}
 C_{\rm v} &
:=
 \int_{\cal D} \td^{3}x \, U,
\\
 C_{\rm m} &
:=
 \int_{\cal D} \td^{3}x \, \psi,
\\
 C_{\rm p} &
:=
 \int_{\cal D} \td^{3}x \, P.
\end{align}

\subsection{Artificial dynamics for relaxation}
\label{subsec:SA}

In this paper we adopt an artificial dynamics defined by the following double  bracket:\cite{Flierl-Morrison-11, Morrison-17} 
\begin{equation}
 (( F, G ))
:=
 \int_{\cal D} \td^{3}x^{\pr}
 \int_{\cal D} \td^{3}x^{\pr\pr} \,
 \{ F, u^{i}(\vx^{\pr}) \}
 K_{ij}(\vx^{\pr}, \vx^{\pr\pr})
 \{ u^{j}(\vx^{\pr\pr}), G \}\,,
\label{eq:double-bracket}
\end{equation}
defined for any functionals $F[\vu]$ and $G[\vu]$, where   $i,j=1,2,3$ and 
$( K_{ij} )$ is a symmetric kernel with a definite sign.
The curly bracket in Eq.~(\ref{eq:double-bracket}) is in general any  Poisson bracket of the form of \eqref{eq:Poisson-bracket}. 
When we take $( K_{ij} )$ as positive definite, then the energy of the system decreases monotonically by the artificial dynamics
\begin{equation}
 \frac{\pd \vu}{\pd t} 
= 
 (( \vu, H )).
\label{eq:SA}
\end{equation}

For the case at hand, it is convenient to  introduce the following artificial convection fields:  
\begin{align}
 \tilde{\vphi}(\vx)
&:=
 - \int_{\cal D} \td^{3} x^{\pr} \,
% \left(
%     K_{11}(\vx, \vx^{\pr}) f^{U}(\vx^{\pr})
%   + K_{12}(\vx, \vx^{\pr}) f^{\psi}(\vx^{\pr})
%  + K_{13}(\vx, \vx^{\pr}) f^{P}(\vx^{\pr})
% \right),
 K_{1i}(\vx, \vx^{\pr}) f^{i}(\vx^{\pr}),
\\
 \tilde{J}(\vx)
&:=
 - \int_{\cal D} \td^{3} x^{\pr} \,
% \left(
%     K_{21}(\vx, \vx^{\pr}) f^{U}(\vx^{\pr})
%   + K_{22}(\vx, \vx^{\pr}) f^{\psi}(\vx^{\pr})
%   + K_{23}(\vx, \vx^{\pr}) f^{P}(\vx^{\pr})
% \right),
 K_{2i}(\vx, \vx^{\pr}) f^{i}(\vx^{\pr}),
\\
 \tilde{h}(\vx)
&:=
 - \int_{\cal D} \td^{3} x^{\pr} \,
% \left(
%     K_{31}(\vx, \vx^{\pr}) f^{U}(\vx^{\pr})
%   + K_{32}(\vx, \vx^{\pr}) f^{\psi}(\vx^{\pr})
%   + K_{33}(\vx, \vx^{\pr}) f^{P}(\vx^{\pr})
% \right),
 K_{3i}(\vx, \vx^{\pr}) f^{i}(\vx^{\pr})  %, 
\end{align}
%with the summation over $i = 1, 2, 3$ as usual, 
in terms of which the artificial evolution of Eq.~(\ref{eq:SA}) can be  written compactly as
\begin{alignat}{2}
  \frac{\pd U}{\pd t} 
&= 
 [ U, \tilde{\vphi} ] + [ \psi, \tilde{J} ] 
 - \veps \frac{\pd \tilde{J}}{\pd \zeta}
 + [ P, \tilde{h} ]
&\,=:\,&
%\tilde{f}^{U},
\tilde{f}^{1},
\label{eq:SA-vorticity-eq}
\\
  \frac{\pd \psi}{\pd t} 
&=
 [ \psi, \tilde{\vphi} ] - \veps \frac{\pd \tilde{\vphi}}{\pd \zeta}
&\,=:\,&
%\tilde{f}^{\psi},
\tilde{f}^{2},
\label{eq:SA-Ohm-law}
\\ 
 \frac{\pd P}{\pd t} 
&= 
 [ P, \tilde{\vphi} ]
&\,=:\,&
%\tilde{f}^{P}.
\tilde{f}^{3}.
\label{eq:SA-pressure-eq}
\end{alignat}
As we see, the convection fields $\vphi$, $J$ and $h$ of the original
evolution equations (\ref{eq:vorticity-eq}), (\ref{eq:Ohm-law}) and 
(\ref{eq:pressure-eq}) are replaced by the artificial convection fields 
$\tilde{\vphi}$, $\tilde{J}$ and $\tilde{h}$, respectively.
The right-hand sides 
%$\tilde{f}^{U}$, $\tilde{f}^{\psi}$ and $\tilde{f}^{P}$ 
$\tilde{f}^{1}$, $\tilde{f}^{2}$ and $\tilde{f}^{3}$ 
are again defined for later use.

Finally we explain the boundary conditions.  We will  use a Fourier mode expansion in the poloidal
and toroidal directions.  For the poloidal mode number $m=0$ components, the radial derivatives are set to be zero at $r=0$,   
for the other $m$ components, their values are set to be zero.  At the plasma edge, all the Fourier components are set to be zero.

\section{Numerical schemes and convergence criterion}
\label{sec:scheme-convergence}

\subsection{Numerical schemes}
\label{subsec:schemes}

In addition to the Fourier mode expansion in $\theta$ and $\zeta$, the  numerical scheme we adopt  for solving Eqs.~(\ref{eq:SA-vorticity-eq})--(\ref{eq:SA-pressure-eq}) employs a  second-order finite difference method in $r$.   Here we only consider   axisymmetric configurations; consequently,  the mode number $n$ for $\zeta$ only takes the value $0$.
 The second-order implicit Runge-Kutta method is used for the time
stepping, which is time-reversal symmetric and symplectic.
The implicit equation associated with this procedure  is solved by the Newton--Raphson method that uses 
the generalized minimum residual (GMRES) method\cite{GMRES} implemented in
the Jacobian-free form.

We assume that the symmetric kernel $( K_{ij} )$ in Eq.~(\ref{eq:double-bracket}) is diagonal,  as in 
Ref.~\onlinecite{Furukawa-17}, and we  choose each diagonal term to have the form  $\alpha_{i} g(\vx^{\pr}, \vx^{\pr\pr})$,
where $\alpha_{i}$ is a constant and   $g(\vx^{\pr}, \vx^{\pr\pr})$ is the  three-dimensional Green's function defined through the Poisson equation
\begin{equation}
 \bigtriangleup g(\vx, \vx^{\pr\pr}) = - \delta^{3}( \vx - \vx^{\pr\pr} ).
\end{equation}
Here the three-dimensional Laplacian is written as $\bigtriangleup$ and  $\delta^{3}( \vx - \vx^{\pr\pr} )$ is the three-dimensional Dirac  delta function.

\subsection{Convergence criterion}
\label{subsec:convergence}

The code used for the  present calculations evolves  the Fourier components of $U$, $\psi$ and $P$.  Thus the right-hand side of the evolution equations (\ref{eq:SA-vorticity-eq})--(\ref{eq:SA-pressure-eq})  is also Fourier decomposed. We write the Fourier
coefficients of $((\vu, H))$, i.e., 
%$\tilde{f}^{U}$, $\tilde{f}^{\psi}$ and $\tilde{f}^{P}$,  
%$\tilde{f}^{i} \,\, (i = 1, 2, 3)$
$\tilde{f}^{i}$
as
%$\tilde{f}^{U}_{mn}$, $\tilde{f}^{\psi}_{mn}$ and  $\tilde{f}^{P}_{mn}$, 
$\tilde{f}^{i}_{mn}$, 
where $m$ and $n$ denote the poloidal and
 toroidal mode numbers, respectively. We also write the Fourier
 coefficients of 
%$f^{U}$, $f^{\psi}$ and $f^{P}$
%$f^{i} \,\,  (i = 1, 2, 3)$
$f^{i}$
 of the original evolution equations (\ref{eq:vorticity-eq})--(\ref{eq:pressure-eq}) as
%$f^{U}_{mn}$, $f^{\psi}_{mn}$ and $f^{P}_{mn}$, 
$f^{i}_{mn}$.
%, respectively. 
Then the convergence criterion is set such that the maximum absolute
values among 
%$f^{U}_{mn}$, $f^{\psi}_{mn}$, $f^{P}_{mn}$,
%$\tilde{f}^{U}_{mn}$, $\tilde{f}^{\psi}_{mn}$ and $\tilde{f}^{P}_{mn}$
$f^{i}_{mn}$ and $\tilde{f}^{i}_{mn}$ 
become smaller than a threshold value.  We took $10^{-6}$, or smaller in
some cases, 
 as the 
threshold value in the numerical results in this paper.

\section{Numerical results}
\label{sec:results}

\subsection{Axisymmetric tokamak equilibrium without rotation}
\label{subsec:tokamak-equilibrium-norotation}

\begin{figure}[h]
 \centering
 \includegraphics[width=0.75\textwidth]{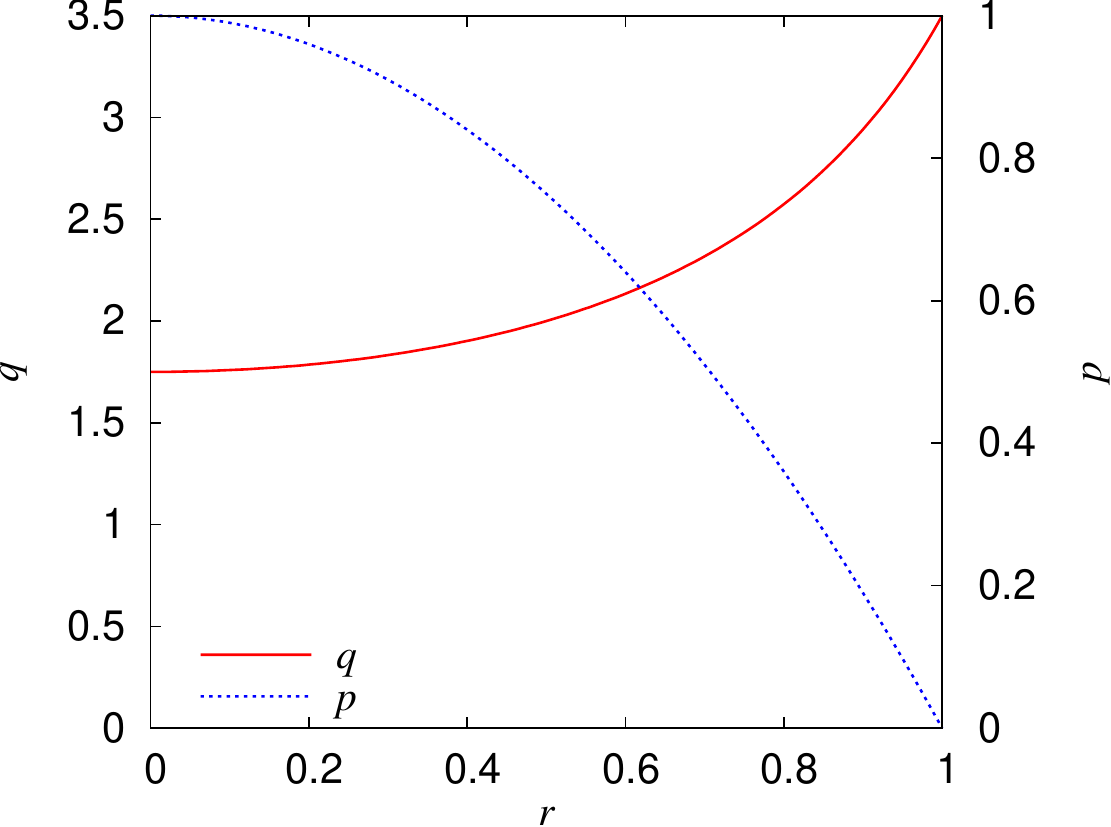}
 \caption{
 \label{fig:r-q-p-initial}
The initial safety factor $q$ and the pressure $p$ profiles.
 }
\end{figure}

Let us turn to our first example, the calculation of  large-aspect-ratio,  circular-cross-section tokamak
equilibria using SA.  The initial conditions are chosen to be cylindrically symmetric with the  profiles of the 
safety factor $q$ and the pressure $p$  as  plotted  in
Fig.~\ref{fig:r-q-p-initial}.  We take $P = \betao p$ and consider three
cases with  
%$\betao = 10^{-3}$, $5 \times 10^{-3}$ and $10^{-2}$.   
$\betao = 0.1$ \%, $0.5$ \% and $1$ \%.   
The rotation velocity is chosen to be zero and, consistently,  the
stream function  $\vphi \equiv 0$.  In the next  subsection, we will consider the case of finite poloidal rotation.   Throughout, the  inverse aspect ratio is set to be  $\veps =  {1}/{10}$.

\begin{figure}[h]
 \centering
 \includegraphics[width=0.75\textwidth]{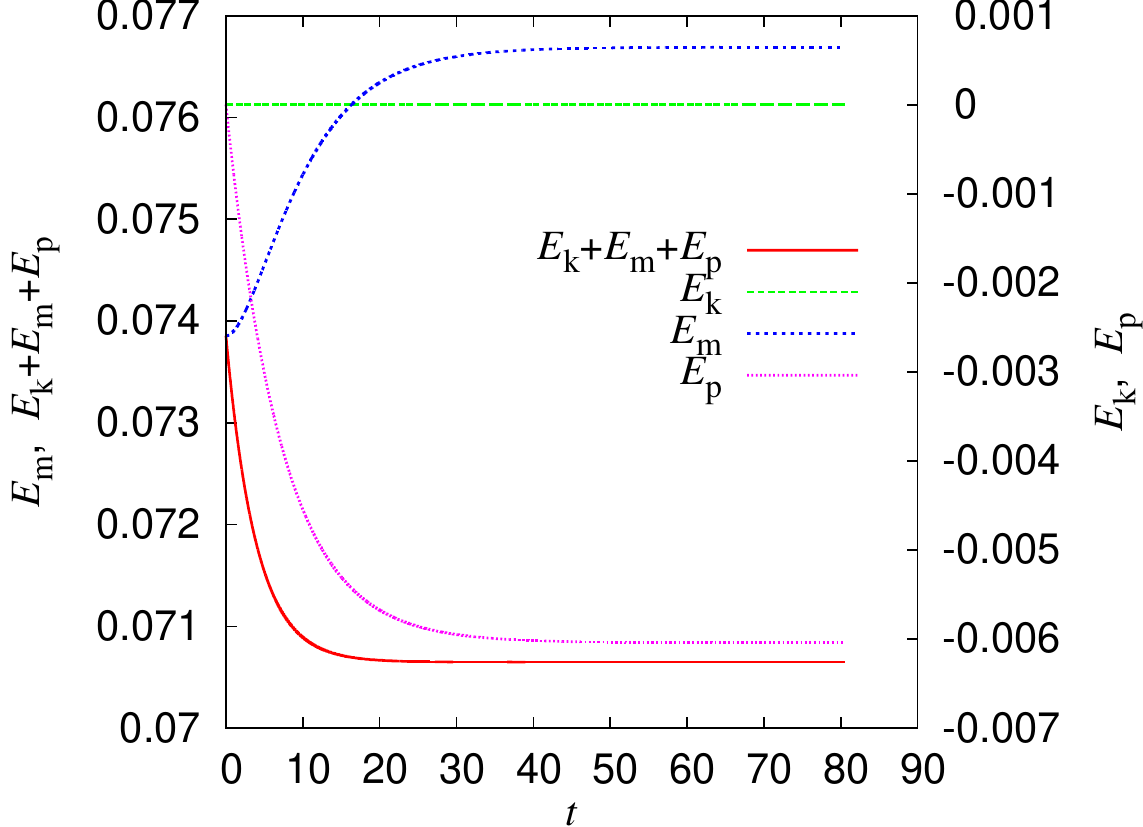}
 \caption{
 \label{fig:t-E-betao1m2}
Time evolution of the kinetic energy $E_{\rm k}$, 
the magnetic energy $E_{\rm m}$,
the internal energy $E_{\rm p}$ and 
the total energy $E_{\rm k} + E_{\rm m} + E_{\rm p}$.
The magnetic energy increases as the time proceeds, while the internal energy decreases more, resulting in the total energy decreasing  monotonically. 
 }
\end{figure}

The time evolution of the total and various components of the energy are shown in Fig.~\ref{fig:t-E-betao1m2}
for the case of 
%$\betao = 10^{-2}$.   
$\betao = 1$ \%.   
The kinetic, magnetic,  and internal energies are denoted by $E_{\rm k}$, $E_{\rm m}$ and $E_{\rm p}$, respectively. The magnetic energy increases as the time proceeds while the internal energy decreases with the 
total energy  decreasing  monotonically.   Note that the kinetic energy remains  zero, a consequence of 
$((U, H))$ being zero for the chosen initial condition,  resulting in  $U$ remaining zero.  
Note,  that the time evolution for the cases 
%$\betao = 10^{-3}$ and   $5 \times 10^{-3}$ 
$\betao = 0.1$ \% and   $0.5$ \%
is similar to those of Fig.~\ref{fig:t-E-betao1m2}.   Also note the Casimir invariants $C_{\rm v}$, $C_{\rm m}$,  and $C_{\rm p}$ are well conserved in the simulation.

\begin{figure}[h]
 \centering
% \subfigure[$\betao=10^{-3}$]{%
 \subfigure[$\betao=0.1$ \%]{%
 \includegraphics[width=0.3\textwidth]{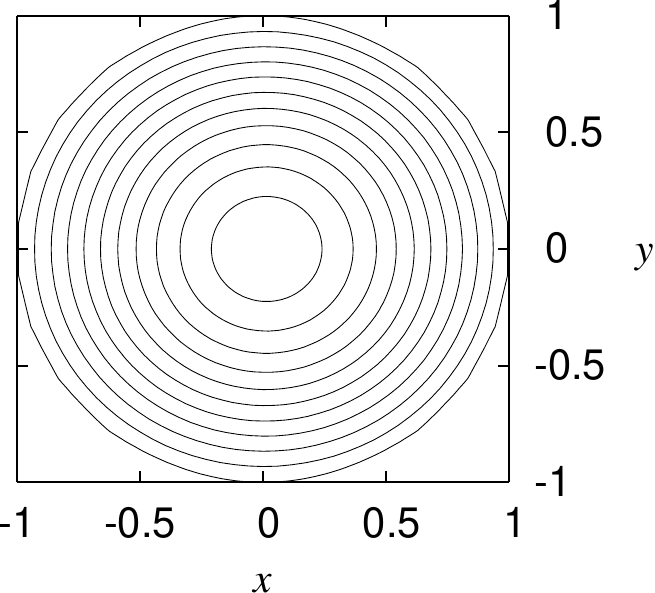} \label{subfig:betao1m3-psi-contour} }
% \subfigure[$\betao=5 \times 10^{-3}$]{%
 \subfigure[$\betao=0.5$ \%]{%
 \includegraphics[width=0.3\textwidth]{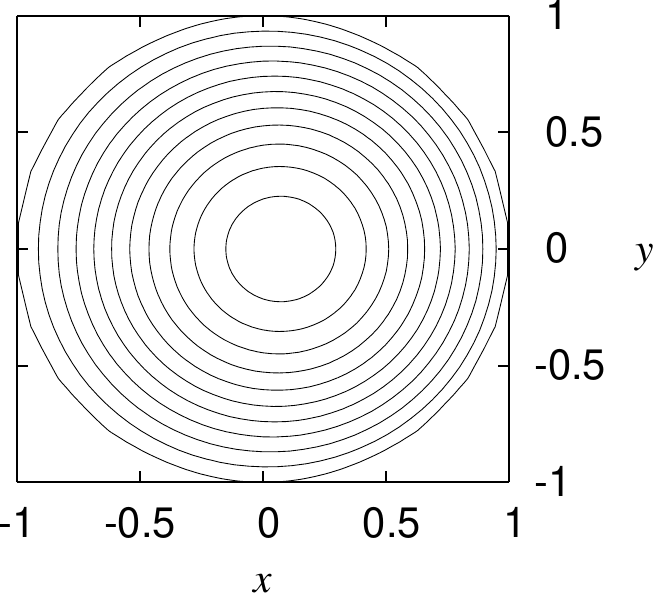} \label{subfig:betao5m3-psi-contour} }
% \subfigure[$\betao=10^{-2}$]{%
 \subfigure[$\betao=1$ \%]{%
 \includegraphics[width=0.3\textwidth]{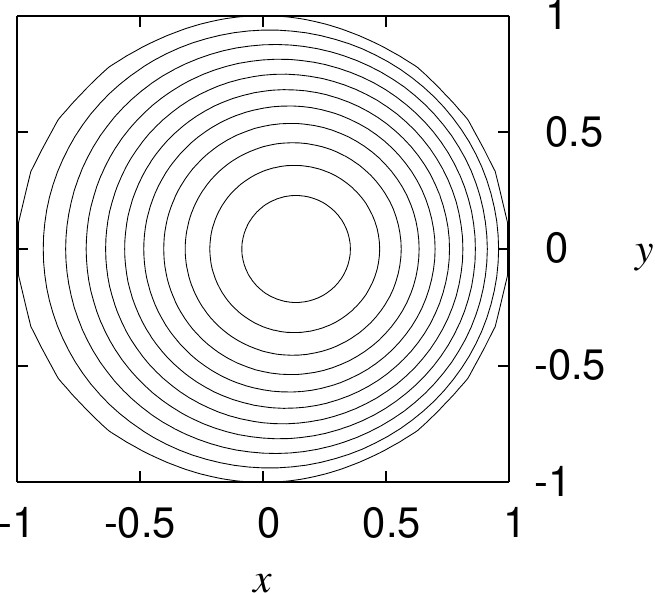} \label{subfig:betao1m2-psi-contour} }
 \caption{
 \label{fig:betao-psi-contour}
Contour plots of $\psi$ for 
%(a) $\betao = 10^{-3}$, (b) $5\times 10^{-3}$ and (c) $10^{-2}$.  
(a) $\betao = 0.1$ \%, (b) $0.5$ \% and (c) $1$ \%.  
The horizontal axis $x$ denotes the   distance along the major radius direction measured from the center of the plasma $R_{0}$.  The vertical axis  $y$ is the height from the midplane.
 }
\end{figure}

Figure~\ref{fig:betao-psi-contour} shows  contour plots of $\psi$ for 
the  equilibria obtained in our three cases.  
The horizontal axis  $x$ is the  distance along the major radius measured  from the plasma center $R_{0}$.  The vertical axis   $y$ is the height from the midplane.  Observe,  the Shafranov shift increases as $\betao$ is increased.

\begin{figure}[h]
 \centering
 \includegraphics[width=0.75\textwidth]{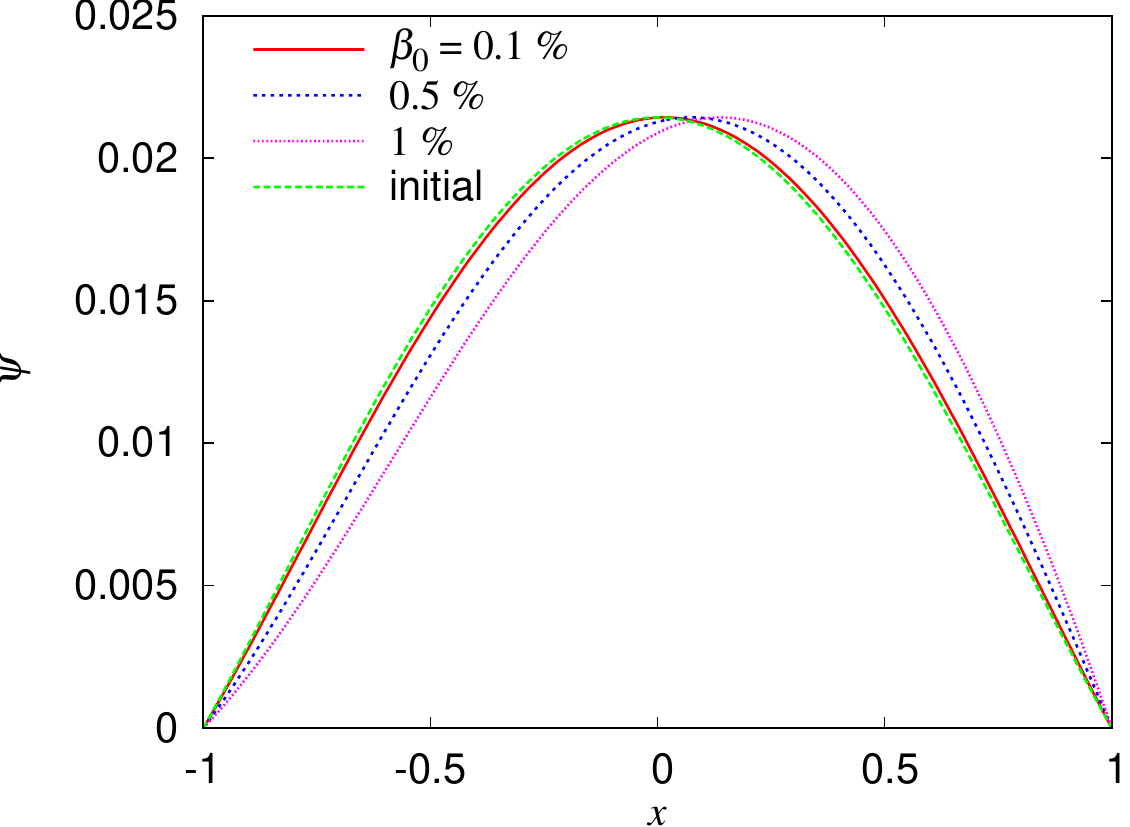}
 \caption{
 \label{fig:x-psi-final}
The $\psi$ profiles  on the midplane for three  equilibria obtained by simulated annealing, along with the 
 initial $\psi$ profile common to each case.
 }
\end{figure}

Figure \ref{fig:x-psi-final} shows plots of  the $\psi$ profiles on the midplane of the three obtained
equilibria, along with the  initial $\psi$ profile common to each case.    As $\betao$ is increased, the peak of $\psi$ 
moves outward.  This is due to the increase of the $m=\pm 1$ and $n=0$ components of $\psi$.  For these numerical results, 
the $m=\pm 2$ and $n=0$ components are very small for the inverse aspect ratio $\veps = {1}/{10}$.    The pressure profiles also have the similar outward shift of their peaks.

\begin{figure}[h]
 \centering
 \includegraphics[width=0.75\textwidth]{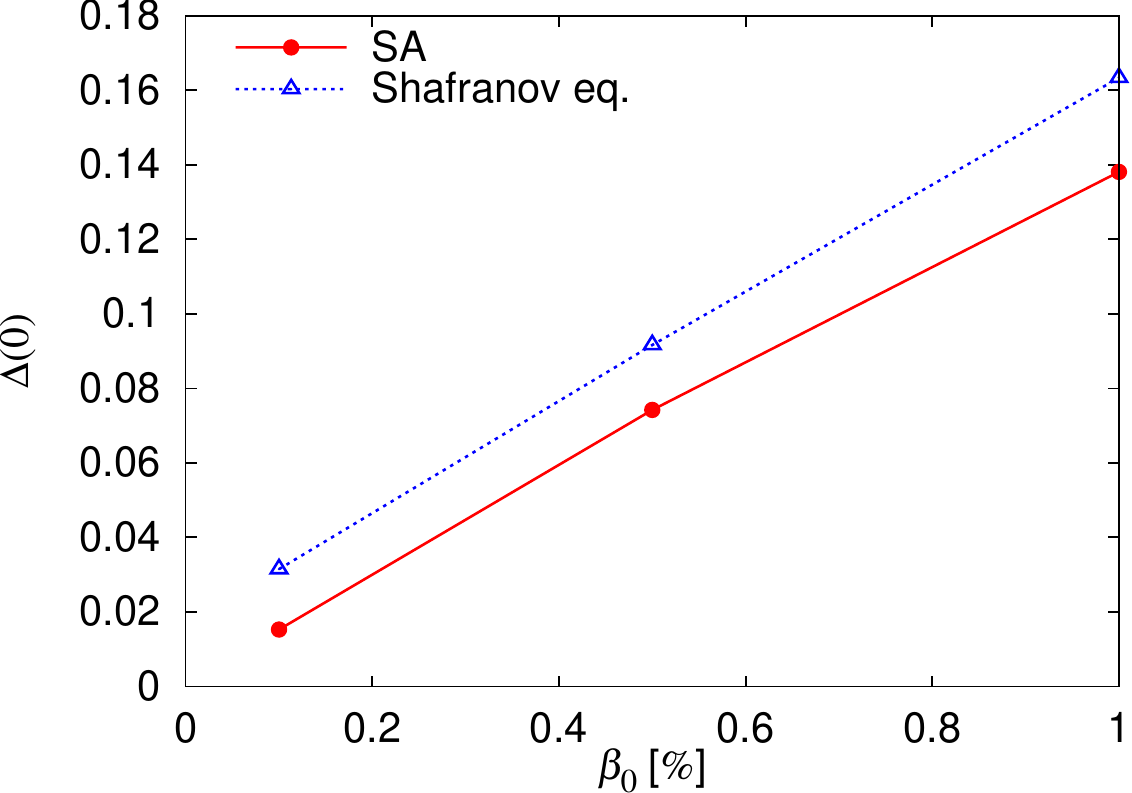}
 \caption{
 \label{fig:betao-Delta0}
The Shafranov shift $\Delta(0)$ vs.\  $\betao$.  Here 
``SA'' denotes the simulated annealing result, while  ``Shafranov eq.'' denotes
the analytic solution obtained by   large-aspect-ratio expansion.
 }
\end{figure}

From the peaks of the $\psi$ profiles, we evaluated the Shafranov shift $\Delta(0)$, which  is shown in Fig.~\ref{fig:betao-Delta0}.  The result of our calculations denoted by ``SA'' are compared with the  Shafranov shift obtained from the  analytic solution using the large-aspect-ratio expansion\cite{Shafranov-66},    denoted by  ``Shafranov eq.''.  In the latter, the Shafranov shift remains at
$\betao = 0$ since the toroidicity remains even at zero beta, however, 
the Shafranov shift seems to be zero at $\betao = 0$ for SA.  This is because the toroidal effect fully disappears at zero
beta in  high-beta reduced MHD.

\begin{figure}[h]
 \centering
 \includegraphics[width=0.75\textwidth]{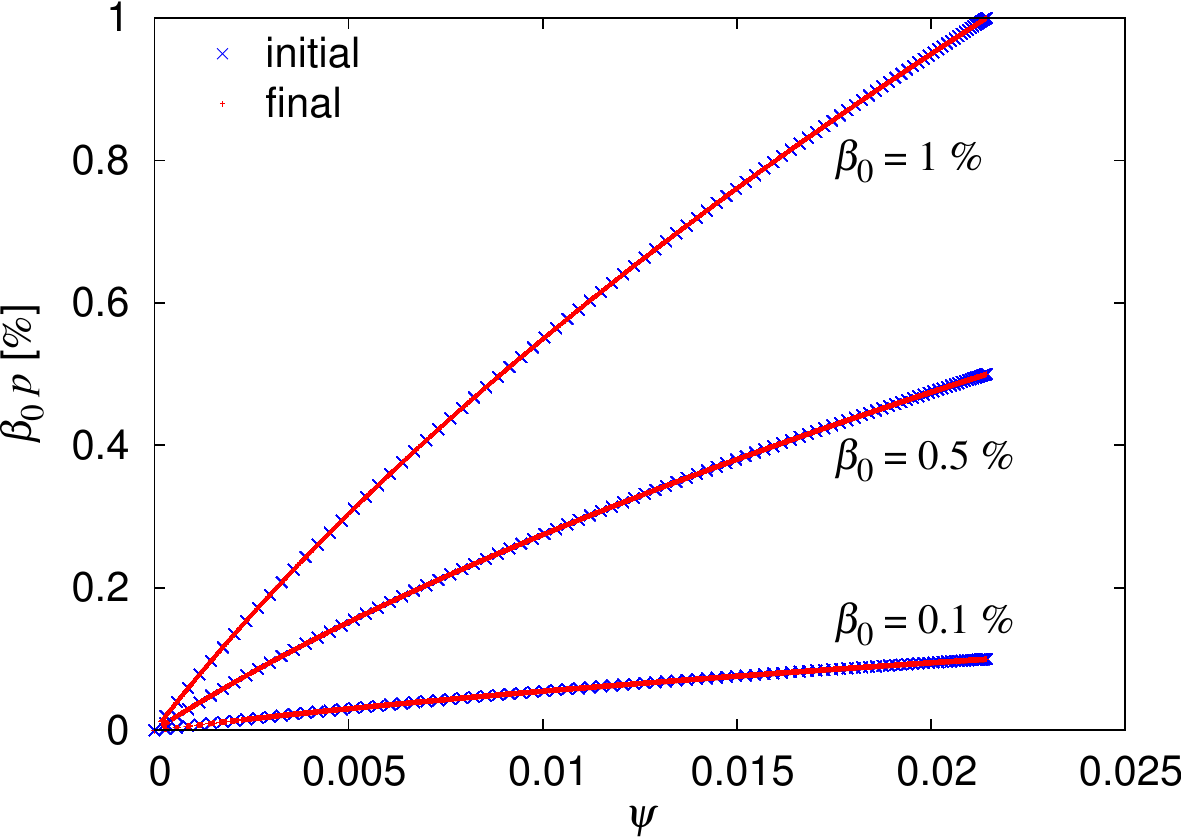}
 \caption{
 \label{fig:psi-p-betao}
The relation between $\psi$ and $P= \betao p$ is shown in a scatter plot
 using $\psi$ and $P$ at $10^{4}$ points on the poloidal cross section.
For each $\betao$, the initial relation between $\psi$ and $P$
 is retained.
 }
\end{figure}

We also note that $\psi$ and $P$ are convected by the same convection
field $\tilde{\vphi}$ in the two-dimensional case.  Therefore, the initial
relation between $\psi$ and $P$ is retained, i.e., if we give an initial
condition in a form $P = P(\psi)$, the same relation holds in the
obtained equilibrium.  This is shown in Fig.~\ref{fig:psi-p-betao}.
We observe that the initial relation between $\psi$ and $P$ is retained
for each $\betao$.
This means, in a sense, that we can specify the pressure profile of the
two-dimensional equilibrium to be obtained, as is usual when  solving  the
Grad-Shafranov equation. 

With SA the total poloidal flux is conserved because  of the Dirichlet boundary condition at the plasma edge.
Therefore the maximum value of $\psi$ does not change from the initial condition,  as we see in Fig.~\ref{fig:psi-p-betao} as well as in
Fig.~\ref{fig:x-psi-final}.  However, the safety factor profile changes
from the initial condition.   In tokamak equilibrium calculations, it may be necessary to
control the safety factor profile.    Such local profile control using  SA will be a future issue.

\subsection{Axisymmetric tokamak equilibrium with poloidal rotation}
\label{subsec:tokamak-equilibrium-poloidal-rotation}

\begin{figure}[h]
 \centering
 \includegraphics[width=0.75\textwidth]{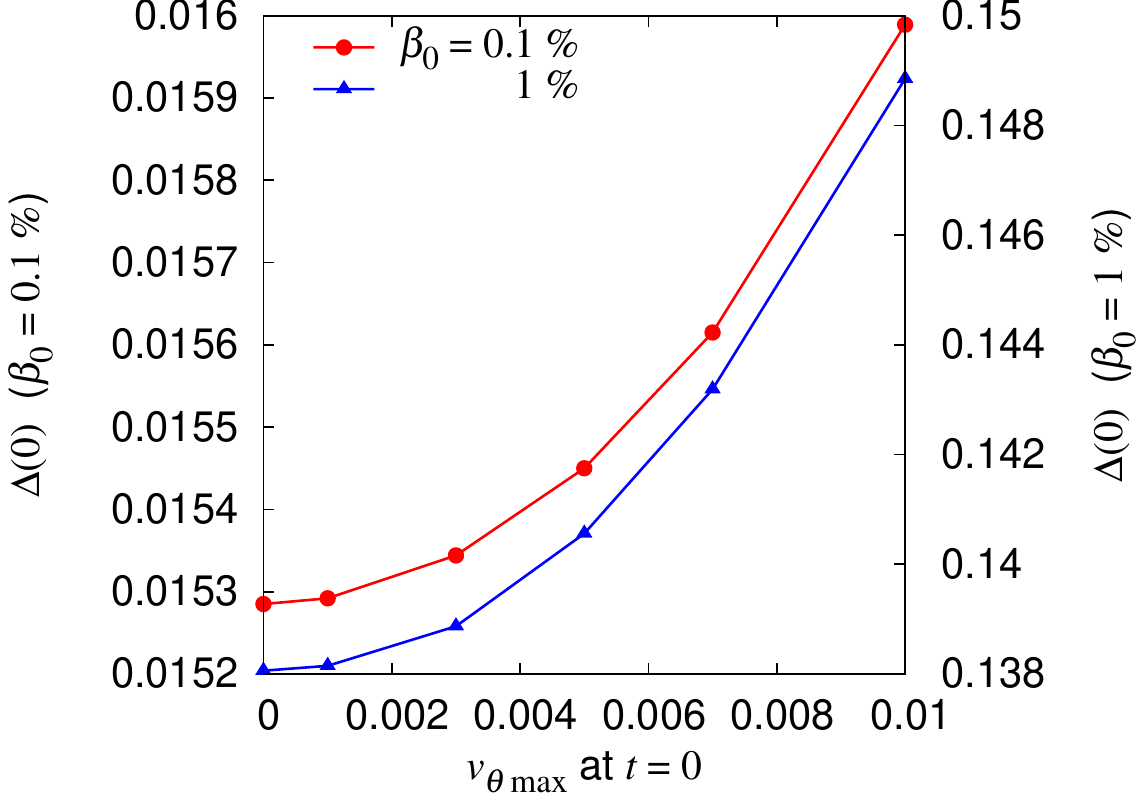}
 \caption{
 \label{fig:vth-Del0-betao}
The effect of poloidal rotation on the Shafranov shift for  
%$\betao = 10^{-3}$ and $10^{-2}$. 
$\betao = 0.1$ \% and $1$ \%. 
The Shafranov shift increases quadratically in  $v_{\theta {\rm max}}$.
 }
\end{figure}

One of the advantages of  SA is that it can handle various types of plasma rotations.  In the present model, the rotation is in the poloidal
direction.  Because the flow is incompressible, there is no difficulty
in calculating such  equilibria.  We performed SA starting from the same
initial conditions in the previous subsection for $\psi$ and $P$, but
now we choose a  $\vphi$ that 
gives $v_{\theta} = 4 v_{\theta {\rm max}} r ( 1 - r )$. Therefore
$v_{\theta}$ has its maximum value $v_{\theta {\rm max}}$ at $r=1/2$
when $t=0$. 
The rotation profile can change as  time proceeds. Indeed, $\vphi$ must be a function of $\psi$ in the equilibrium according to the Ohm's law (\ref{eq:Ohm-law}) with axisymmetry.   The flux surfaces shift outward in the major radius direction at finite
beta, and thus $\vphi$ must also change from the initial condition to meet the equilibrium condition.  
Figure~\ref{fig:vth-Del0-betao} shows that  the Shafranov shift of the magnetic axis $\Delta(0)$  increases approximately quadratically  in the poloidal rotation speed.   This is due to the dynamic pressure of the flow.   Here the poloidal rotation speed is normalized by the toroidal Alfv\'en velocity, but if we normalize it by the poloidal Alfv\'en velocity at the plasma edge, $v_{\theta} = 10^{-2}$ corresponds to 3.5 \% of the poloidal Alfv\'en velocity.   The increase of the Shafranov shift is about 5 -- 7 \% compared to the non-rotating equilibrium at $v_{\theta {\rm max}} = 10^{-2}$, however, the  flux surface shape does not change much.

Here we explain a mapping procedure between an equilibrium without rotation
and poloidally rotating equilibria for the high-beta reduced MHD,
and then show that the quadratic dependence on the poloidal rotation
speed can be explained by the map.
The map is an extension of a map first derived for the low-beta reduced MHD
in Ref.~\onlinecite{Morrison-86} and a special case of that of
Refs.~\onlinecite{Throumoulopoulos-99, Andreussi-12}. 
The equilibrium equations including the poloidal rotation are given by 
$f^{i} = 0 \,\, (i = 1, 2, 3)$.
From the equation for $i=2$, or $[ \psi, \vphi ] = 0$, 
we obtain $\vphi = G(\psi)$ where $G$ is an arbitrary function.
Also from the equation for $i=3$, or $[ P, \vphi ] = 0$, 
we obtain $P = K(\vphi) = K \circ G(\psi) =: L(\psi)$ where $K$ is an
arbitrary function.  Noting that
$J = \bigtriangleup_{\perp} \psi$ 
and
\begin{equation}
 U = \bigtriangleup_{\perp} \vphi
=
  G^{\pr}(\psi) \bigtriangleup_{\perp} \psi
+ G^{\pr\pr}(\psi) | \nab_{\perp} \psi |^{2},
\end{equation}
$f^{1} = 0$ reads
\begin{equation}
 [ G^{\pr}(\psi) \bigtriangleup_{\perp} \psi
+ G^{\pr\pr}(\psi) | \nab_{\perp} \psi |^{2}
 , \psi ]
- [ \bigtriangleup_{\perp} \psi, \psi ]
- [ h L^{\pr}(\psi), \psi ]
= 0.
\end{equation}
Therefore we obtain\cite{Marsden-84}
\begin{equation}
 ( 1 - G^{\pr 2}(\psi) ) \bigtriangleup_{\perp} \psi
- G^{\pr\pr}(\psi) | \nab_{\perp} \psi |^{2} 
+ h L^{\pr}(\psi) + F(\psi)
= 0
\label{eq:GSeq-rot}
\end{equation}
where $F(\psi)$ is an arbitrary function.
This is an extended Grad-Shafranov equation for a poloidally rotating
equilibrium.
By setting $G(\psi) \equiv 0$, we obtain the Grad-Shafranov equation without
a rotation as 
\begin{equation}
 \bigtriangleup_{\perp} \psi
+ h L^{\pr}(\psi) + F(\psi)
= 0.
\label{eq:GSeq-norot}
\end{equation}
Now, if we define
\begin{equation}
 \chi = X(\psi)
:= 
 \int^{\psi} \sqrt{ 1 - G^{\pr 2}(\psi^{\pr}) } \, \td \psi^{\pr},
\end{equation}
the equilibrium equation can be rewritten as
\begin{equation}
 \bigtriangleup_{\perp} \chi
+ h \bar{L}^{\pr}(\chi) 
+ \bar{F}(\chi)
= 0,
\label{eq:GSeq-rot-map}
\end{equation}
where
\begin{align}
 \bar{L}^{\pr}(\chi)
&:=
 \frac{ L^{\pr}(X^{-1}(\chi)) }
      { \sqrt{ 1 - G^{\pr 2}(X^{-1}(\chi)) } },
\\
 \bar{F}(\chi)
&:=
 \frac{ F(X^{-1}(\chi)) }
      { \sqrt{ 1 - G^{\pr 2}(X^{-1}(\chi)) } }.
\end{align}
Note that $X^{-1}(\chi) = \psi$.
Equation~(\ref{eq:GSeq-rot-map}) has the same form as
Eq.~(\ref{eq:GSeq-norot}) for an equilibrium without a rotation. 
Therefore an equilibrium without a rotation can be mapped to equilibria
with poloidal rotations.  
We see that the source terms of Eq.~(\ref{eq:GSeq-rot-map}) are changed by
the factor $1 / \sqrt{ 1 - G^{\pr 2}(\psi) }$.
Here $G^{\pr}(\psi)$ expresses the poloidal rotation velocity normalized by
the poloidal Alfv\'en velocity.
When $| G^{\pr 2}(\psi)| \ll 1$,
the factor can be approximated as
$1 / \sqrt{ 1 - G^{\pr 2}(\psi) } \simeq 1 + \frac{1}{2} G^{\pr 2}(\psi)$
that indicates effective increase of the source term
such as the pressure gradient quadratically in the poloidal rotation
speed.  This explains the quadratic dependence of the
Shafranov shift on the poloidal rotation speed. We also point out that any equilibria without rotation can be identidfied with one with rotation by a choice of $G$.  This, of course, includes the equilibria obtained here by SA.

\subsection{Heliotron equilibrium averaged over toroidal direction}
\label{subsec:heliotron-equilibrium}

For our next example  we consider heliotron equilibria  averaged over the toroidal direction.
As mentioned at the end of Sec.~\ref{subsec:high-beta-RMHD}, the reduced MHD equations (\ref{eq:vorticity-eq})--(\ref{eq:pressure-eq})  can be used for stellarators under an appropriate replacement of variables.
Here, we further assume that the dependence of the equilibrium quantities on $\zeta$, the toroidal angle for the long wavelength
structure, vanishes.  Then the $\zeta$-derivative terms drop out.

The poloidal flux function $\psi$ is replaced by a total poloidal flux function $\Psi := \Psi_{\rm h} + \psi$, 
where  $\Psi_{\rm h}$ is independent on time.  Thus,  $\partial \Psi_{\rm h}/\partial t=0$ and only $\psi$ on the right-hand
side needs to be replaced. The poloidal flux function due to  helical coils,  derived under the cylindrical approximation, is given in  normalized form  by
\begin{equation}
 \Psi_{\rm h}
:=
 - \frac{\biota_{\rm h}(a)}{M} \frac{F(M \veps r)}{M \veps},
 \label{eq:helical-flux}
\end{equation}
where $\biota_{\rm h}(a)$ is a rotational transform  generated by the helical coils, evaluated at the plasma edge,
$M$ is the pitch number, $\veps$ is the inverse aspect ratio,
and
\begin{equation}
 F(M \veps r)
:=
 \frac{\ell}{M \veps r} 
 I_{\ell}(M \veps r) I_{\ell}^{\pr}(M \veps r).
\end{equation}
Here $\ell$ is the pole number, 
$I_{\ell}(z)$ denotes the $\ell$-th order modified Bessel function of
the first kind, and $I_{\ell}^{\pr}(z)$ denotes a $z$ derivative of 
$I_{\ell}(z)$.

The curvature term $\Omega$, derived under cylindrical approximation, is given  in  normalized form  by
\begin{equation}
 \Omega
:= 2 \veps r \cos \theta
 + \veps \biota_{\rm h}(a) 
   \frac{G(M \veps r)}{F^{\pr}(M \veps)},
\label{eq:magnetic-curvature}
\end{equation}
where $F^{\pr}(M \veps)$ denotes the derivative of $F$ with respect to 
$M \veps r$ and is evaluated at $r=1$, and
\begin{equation}
 G(M \veps r)
:=
 \left( I_{\ell}^{\pr}(M \veps r) \right)^{2}
 + \frac{ \ell^{2} + (M \veps r)^{2} }{(M \veps r)^{2}}
   \left( I_{\ell}(M \veps r) \right)^{2}.
\end{equation}
The first and the second terms of $\Omega$ are  the  toroidal magnetic field curvature and the helical curvature, respectively.

We seek to reproduce  simulations of  the Heliotron E device published in Ref.~\onlinecite{Nakamura-93}.  We assume the inverse aspect ratio $\veps =  {1}/{10}$, which is  similar to that of  the Heliotron E device, and that the poloidal cross section has a circular shape.    Also we set the pole number $\ell = 2$ and the pitch number $M = 19$.

For the initial pressure profile we take $P = \betao ( 1 - s )^{2}$ with $s := ( \Psi - \Psi(0) ) / ( \Psi(a) - \Psi(0) )$ being a normalized total poloidal flux function, where $\Psi(0)$ and $\Psi(a)$ are evaluated at the magnetic axis and at the plasma edge, respectively.  This is the same profile  as that of Ref.~\onlinecite{Nakamura-93}.   Note that this pressure profile remains  unchanged during the artificial SA dynamics,  as explained in the last part of Sec.~\ref{subsec:tokamak-equilibrium-norotation}.

It is difficult to completely reproduce the results of  Ref.~\onlinecite{Nakamura-93} for several reasons.
First,  the net plasma current in the toroidal direction on each flux surface cannot be kept zero with the form of  SA used here, while 
in Ref.~\onlinecite{Nakamura-93}  it is set to be zero.   The second reason is that there is  difficulty in controlling the 
local profile  of the rotational transform $\biota$ with SA.  Because of these two reasons, we cannot perfectly match the rotational
transform with that of Ref.~\onlinecite{Nakamura-93}.  A third reason is that the expressions for the helical flux 
$\Psi_{\rm h}$ in Eq.~(\ref{eq:helical-flux}) and the magnetic curvature $\Omega$ in Eq.~(\ref{eq:magnetic-curvature}) are derived under the cylindrical approximation, while those in Ref.~\onlinecite{Nakamura-93} are calculated numerically from helical coil currents in  toroidal geometry.  The last reason is due to  the difference of the shape of the plasma edge.  In our
simulation it is exactly circular, while it is slightly elongated   in Ref.~\onlinecite{Nakamura-93}.  The aspect ratio is  also not 
exactly matched.  Despite of these reasons, we observe reasonable agreement between our SA results  and those of  
Ref.~\onlinecite{Nakamura-93}. 

\begin{figure}[h]
 \centering
 \includegraphics[width=0.5\textwidth]{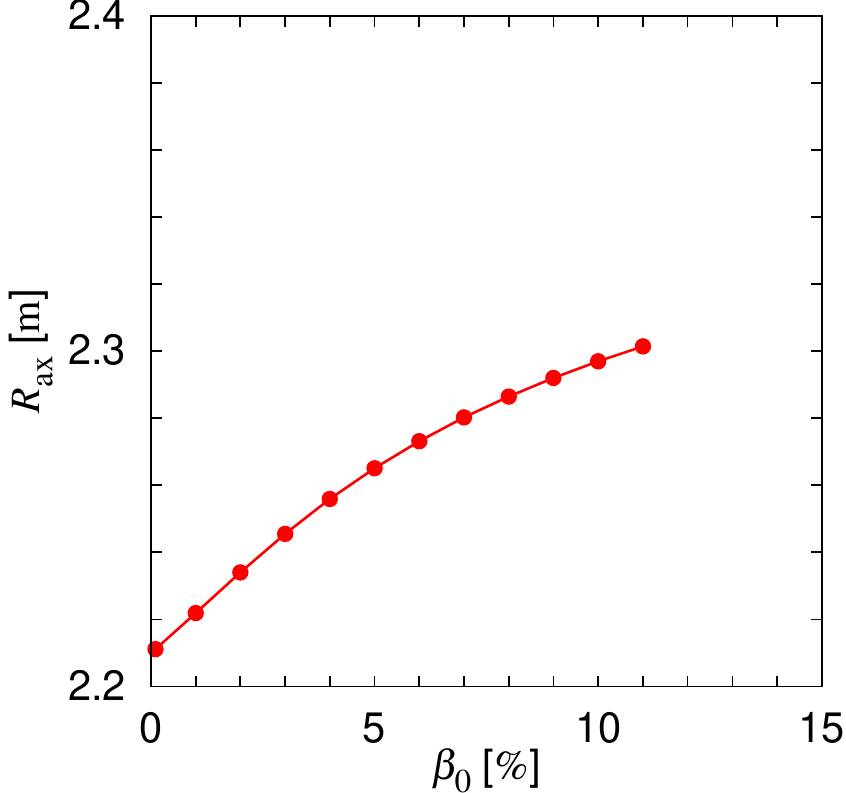}
 \caption{
 \label{fig:betao-Rax-helical}
The magnetic axis position $R_{\rm ax}$  plotted against $\betao$.
 }
\end{figure}

%Figure~\ref{subfig:betao-Rax-helical} 
Figure~\ref{fig:betao-Rax-helical} shows the magnetic axis position  $R_{\rm ax}$ as a function of $\betao$.
Observe that the tendency and the magnitude of the magnetic axis shift agree reasonably with Fig.~2(a) of Ref.~\onlinecite{Nakamura-93}, despite the  differences explained above.

\begin{figure}[h]
 \centering
 \subfigure[$\betao = 3$ \%]{%
 \includegraphics[width=0.32\textwidth]{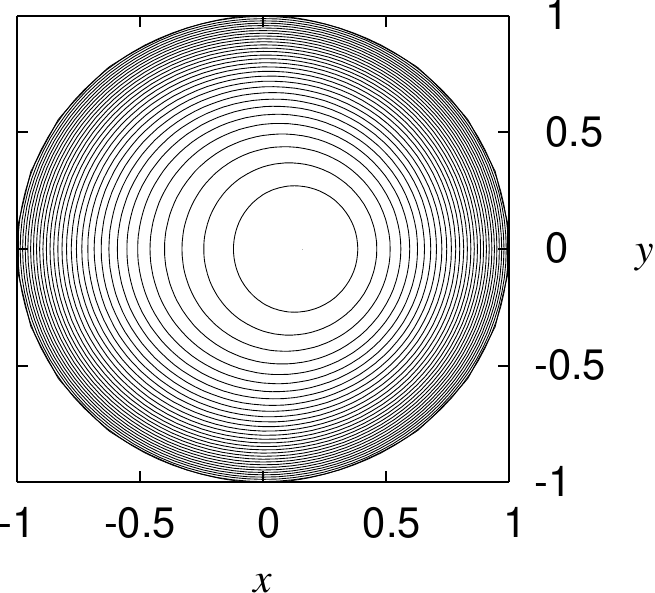} \label{subfig:betao3m2-psi-contour-helical} }
 \subfigure[$\betao = 7$ \%]{%
 \includegraphics[width=0.32\textwidth]{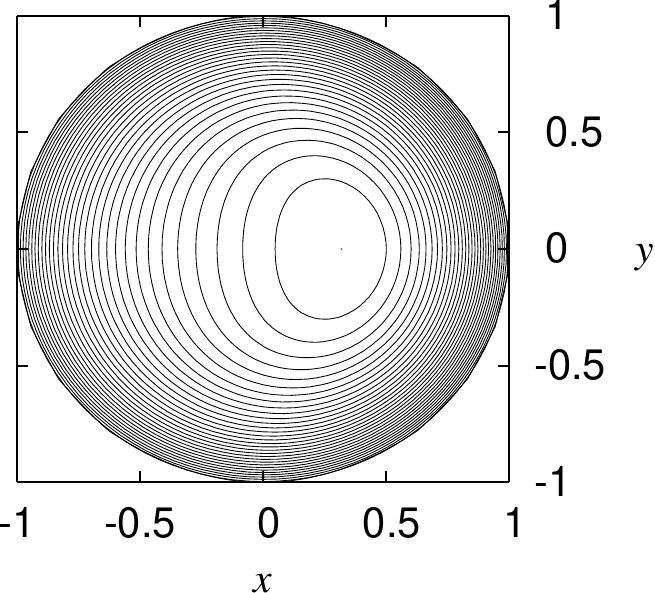} \label{subfig:betao7m2-psi-contour-helical} }
 \subfigure[$\betao = 11$ \%]{%
 \includegraphics[width=0.32\textwidth]{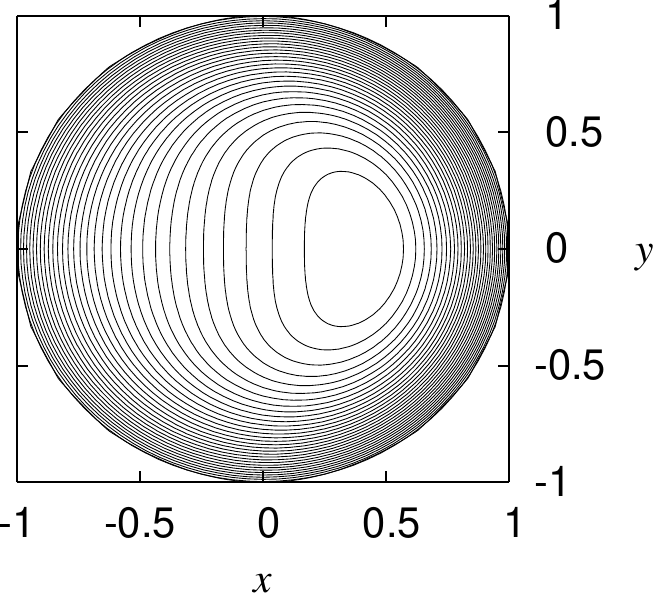} \label{subfig:betao1_1m1-psi-contour-helical} }
 \caption{
 \label{fig:betao-psi-contour-helical}
Contour plots of $\Psi$ for (a) $\betao = 3$ \%,
(b) $\betao = 7$ \% and
(c) $\betao = 11$ \%.
 }
\end{figure}

The shapes of the flux surfaces are shown in Fig.~\ref{fig:betao-psi-contour-helical}.  
A similar figure of  flux surfaces is presented in Fig.~3 of   Ref.~\onlinecite{Nakamura-93} for $\betao = 7$ \%.  
The shapes of the flux surfaces in the central region of the plasma, especially the deformation to the D shape at high beta, 
show behavior similar to  that of  Ref.~\onlinecite{Nakamura-93}.

\begin{figure}[h]
 \centering
 \subfigure[]{%
 \includegraphics[width=0.47\textwidth]{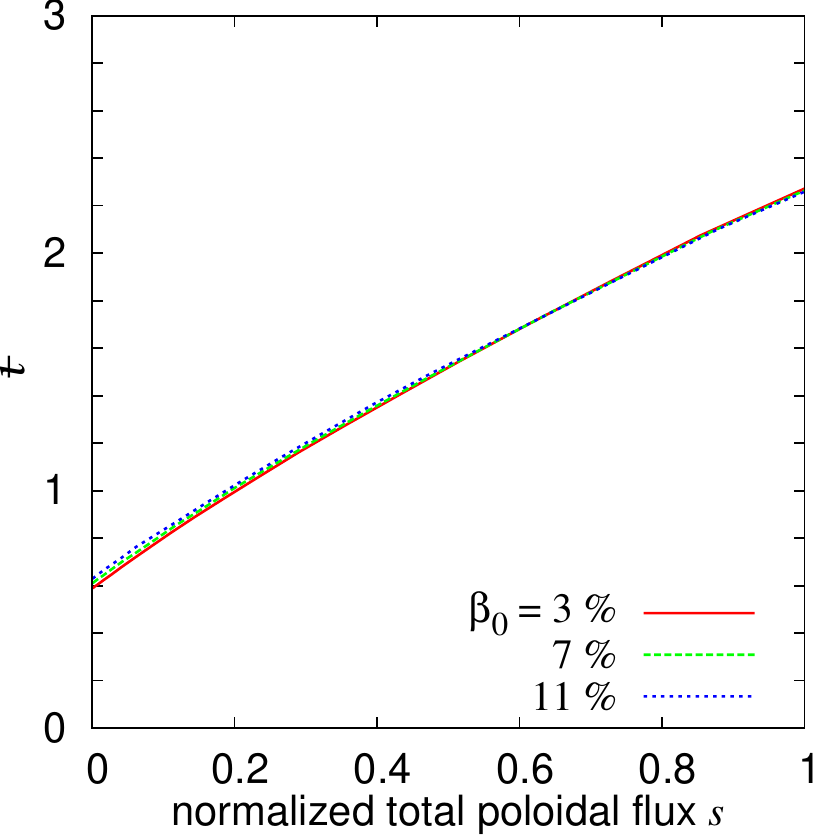} \label{subfig:psin-iota-betao-helical} }
 \subfigure[]{%
 \includegraphics[width=0.47\textwidth]{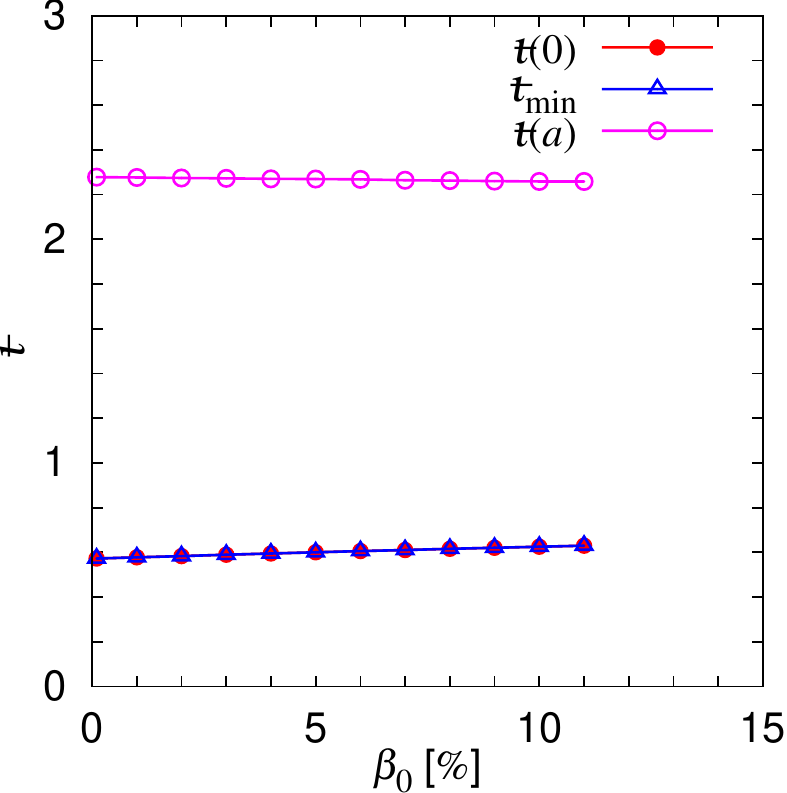} \label{subfig:betao-iota-helical} }
 \caption{
 \label{fig:iota-helical}
(a) The radial profiles of $\biota$ for several values of $\betao$ as
 functions of the normalized total poloidal flux $s$.   (b) The rotational transform at the magnetic axis $\biota(0)$,
at its minimum $\biota_{\rm min}$,  and at the plasma edge $\biota(a)$   plotted against $\betao$.
 }
\end{figure}

Figure \ref{fig:iota-helical} depicts the rotational transform.  Here,
in Fig.~\ref{subfig:psin-iota-betao-helical},  the radial profiles of
$\biota$ for 
%$\betao = 3 \times 10^{-2}$, $7 \times 10^{-2}$,  and $1.1 \times 10^{-1}$
$\betao = 3$ \%, $7$ \%,  and $11$ \%
 are plotted as functions of the 
normalized total poloidal flux $s$.   Also $\biota$  at the magnetic axis $\biota(0)$, at its minimum $\biota_{\rm min}$,  
and at the plasma edge $\biota(a)$ as functions of $\betao$ are shown  in Fig.~\ref{subfig:betao-iota-helical}.
Although it is difficult to identify from the figure,   $\biota(a)$ slightly decreases and $\biota(0)$ slightly increases as 
$\betao$ is increased.  We do not observe a difference between $\biota(0)$ and $\biota_{\rm min}$,  as was  observed in
Ref.~\onlinecite{Nakamura-93}.   The main reason for this difference is that the net current free condition in the equilibria of Ref.~\onlinecite{Nakamura-93} is  not imposed in our equilibria.  The difference between $\biota(0)$ and $\biota_{\rm min}$ arises because of the magnetic shear reversal that is caused by forcing the net zero current.  

In summary, because of some differences in  the equilibrium parameters, e.g.,  the net current, we do not obtain perfect agreement with  Ref.~\onlinecite{Nakamura-93}.  However, we obtain reasonable agreement, especially in the Shafranov shift and the flux surface shapes.

\section{Conclusions}
\label{sec:conclusions}

We have developed the theory of the simulated annealing for calculating
high-beta reduced MHD equilibria in toroidal geometry.
Large-aspect-ratio  circular-cross section tokamak and toroidally
averaged stellarator 
equilibria were successfully calculated.  These equilibria possess the
proper Shafranov shift according to the toroidal effect. 
We obtain reasonable agreement of our results with the previous
studies. Especially for the tokamak equilibrium calculation, we
obtained equilibria including poloidal rotation.  For this case the
Shafranov shift of the magnetic axis increased quadratically  in the
poloidal rotation speed.  
The quadratic dependence is explained by the   mapping
procedure  between non-rotating and poloidally rotating equilibria for 
high-beta reduced MHD.
This achievement highlights one of the
advantages of the simulated annealing method, i.e., it can handle
equilibria with rotation.

% If in two-column mode, this environment will cyhange to single-column format so that long equations can be displayed. 
%\begin{widetext}
%$$\mbox{put long equation here}$$
%\end{widetext}

% If you have acknowledgments, this puts in the proper section head. v
\begin{acknowledgments}
MF was supported by JSPS KAKENHI Grant No. JP15K06647, while PJM was supported by U.S. Dept.\ of Energy Contract \# DE-FG05-80ET-53088.
KI was supported by the budget NIFS17KLTT006
of National Institute for Fusion Science
and JSPS KAKENHI Grant No.15k06651.
\end{acknowledgments}

%\bibliography{/Users/furukawa/texmf/bib/ref.bib}

\end{document}